\documentclass[twocolumn]{aastex631}
\usepackage{amsmath}
\usepackage{amssymb}
\usepackage{bm}
\usepackage{bbm}

\usepackage{float}
\usepackage{graphicx} 
\usepackage{multirow}
\usepackage{hhline}
\usepackage{booktabs}
\usepackage{makecell}
\usepackage{newtxtext,newtxmath}
\usepackage{placeins}
\usepackage{xcolor}
\usepackage{hyperref}

\def\Euclid{\mbox{{\textit{Euclid}}}}
\def\Roman{\mbox{{\textit{Roman}}}}

\newcommand{\dd}{\mathrm{d}}

\definecolor{amber}{rgb}{1.0, 0.49, 0.0}
\definecolor{darkgreen}{rgb}{0.09, 0.45, 0.27}

\begin{document}
\title{Diffusion-based Galaxy Simulations for the Roman High Latitude Survey}

\author[0000-0001-8450-7885]{Diana Scognamiglio}
\affiliation{Jet Propulsion Laboratory, California Institute of Technology, 4800,  Oak Grove Drive - Pasadena, CA 91109, USA}
\affiliation{Duke University, Durham, NC 27708, USA}

\author[0000-0002-2838-2878]{Jake H. Lee}
\affiliation{Jet Propulsion Laboratory, California Institute of Technology, 4800,  Oak Grove Drive - Pasadena, CA 91109, USA}

\author[0000-0002-9378-3424]{Eric Huff}
\affiliation{Jet Propulsion Laboratory, California Institute of Technology, 4800,  Oak Grove Drive - Pasadena, CA 91109, USA}


\author[0000-0003-0220-0009]{Sergi R. Hildebrandt}
\affiliation{Department of Physics, California Institute of Technology, 1200 E. California Boulevard, Pasadena, CA 91125, USA}

\author[0000-0003-2226-5395]{Shoubaneh Hemmati}
\affiliation{IPAC, Caltech, 1200 E. California Blvd., Pasadena, CA 91125}



\begin{abstract}
\noindent
Future weak lensing analyses with the \textit{Nancy Grace Roman} Space Telescope will require highly realistic image simulations to control shear systematics at unprecedented precision. A key limitation of existing approaches is their reliance on analytic light-profile models, which cannot fully capture the complex, non-parametric morphologies revealed by high-resolution observations.
We present a diffusion-based framework for generating realistic galaxy image simulations tailored to the weak lensing requirements of the \textit{Roman} High Latitude Survey. We construct \textit{Roman}-like galaxy images from multi-band JWST/NIRCam observations in the GOODS-S and GOODS-N fields, transforming them into the \textit{Roman} observing regime through point-spread-function matching, pixel-scale conversion, and interloper masking that preserves correlated noise properties. These data are used to train a denoising diffusion probabilistic model to generate multi-band galaxy postage stamps in the \textit{Roman} Y, J, and H filters.
We validate the generated sample against an independent dataset using a consistent photometric pipeline, comparing key galaxy observables including magnitude, size, ellipticity, peak surface brightness, and three-band colors. The generated galaxies reproduce both the marginal distributions and the covariance structure of these properties, with only modest deviations in low-occupancy regions of parameter space.
These results demonstrate that diffusion models provide a scalable and physically motivated alternative to analytic simulations, enabling high-fidelity galaxy populations for \textit{Roman} weak lensing calibration and, more generally, for survey preparation in upcoming cosmological experiments.

\noindent

\end{abstract}



\section{Introduction} \label{sec:intro}
The \textit{Nancy Grace Roman} Space Telescope\footnote{Roman was formerly named the Wide-Field Infrared Survey Telescope (WFIRST).}$^{,}$\footnote{\url{http://roman.gsfc.nasa.gov/}} (hereafter \Roman; \citealt{spergel2015, akeson2019}) will deliver a transformative view of the cosmos with its High Latitude Wide Area Survey (HLWAS; hereafter part of the HLS), a wide-field near-infrared imaging program covering 2,000 square degrees. A central pillar of the HLS is weak gravitational lensing (WL; \citealt{BertSchn01}), which will constrain the growth of large-scale structure and the nature of dark energy. Meeting \Roman’s ambitious cosmological goals requires controlling shear systematics to unprecedented levels, with multiplicative biases $m$ constrained to $m \sim 3.2 \times 10^{-4}$ \citep{Dore2018}. Achieving this accuracy depends critically on the realism of galaxy image simulations, which are used to calibrate shear measurements, test photometric redshifts, and evaluate the impact of blending \citep{Hemmati2019}.

Simulating \Roman-like observations is particularly challenging. The survey will encompass billions of galaxies across a wide range of redshifts, morphologies, and environments, many of which display irregular, clumpy, or multi-component structures only revealed at \textit{Hubble} Space Telescope (HST; \citealt{Lallo_2012}) and \textit{James Webb} Space Telescope (JWST; \citealt{McElwain_2023}) resolution. Traditional pipelines based on analytic light profiles such as Sérsic or double-Sérsic models--offer interpretability and computational control, but they struggle to reproduce the full morphological complexity observed in real galaxies. Moreover, scaling such approaches to the $\sim 10^{9}$ galaxies required for \Roman\ WL calibration remains computationally demanding.

The only end-to-end \Roman\ simulation suite developed to date is that described in \citet{Troxel_2020} and in \citet{OpenUniverse2025}, which constructs synthetic HLS images using \texttt{GalSim}-based galaxy models. In this framework, each galaxy is represented as a combination of bulge, disk, and knot components described by double-Sérsic profiles, with representative spectral energy distributions (SEDs) assigned and evolved across the survey’s passbands. Stars are modeled as point sources, and all objects are subsequently convolved with spatially constant point-spread functions (PSFs) and processed through detector-level effects. This suite has been invaluable for quantifying the impact of wavefront errors and for testing WL analysis pipelines \citep{Hirata_2024, Yamamoto_2024, Berlfein_2025}. Nevertheless, its reliance on analytic light profiles inherently limits the diversity of morphological features that can be reproduced---features such as asymmetric star-forming clumps, tidal debris, or irregular structures that are common among galaxies.

Recent advances in machine learning, and in particular diffusion-based generative models \citep{ho2020DDPM, Smith2022, lizarraga2024}, provide a powerful new avenue to overcome the limitations of analytic galaxy simulations. Diffusion models learn the full high-dimensional distribution of galaxy morphologies directly from imaging data, rather than relying on predefined functional forms. Once trained on high-resolution multi-wavelength datasets, these models can generate synthetic galaxies that capture realistic non-parametric structures while maintaining consistency with the statistical properties of the training population. Crucially, diffusion models can scale efficiently to the massive data volumes required for next-generation WL surveys.

In \citet{Scognamiglio_2025}, we demonstrated the first application of this framework to cosmology by generating realistic \Euclid-like galaxy images. That work showed that DDPMs are capable of producing high-fidelity galaxy populations whose structural and photometric parameter distributions closely match those of validation datasets, confirming their utility for WL shear calibration.

Here, we extend this approach to the \Roman’s HLS. Building on multi-band JWST observations, we train diffusion models to create \Roman-like galaxies across the relevant filters and depths, reproducing structural and photometric properties. Our pipeline complements existing \texttt{GalSim}-based simulations suite: by combining the interpretability of analytic models with the morphological realism and scalability of generative AI, we provide the community with a more flexible and robust toolkit for \Roman\ cosmology. Beyond shear calibration, diffusion-based simulations enable studies of blending, source detection \citep{Hemmati_2022}, study the dependence on galaxy morphology, ultimately enhancing the cosmological and astrophysical return of \Roman\ and its synergy with \Euclid\footnote{\url{https://sci.esa.int/Euclid/}} \citep{laureijs2011euclid} and \textit{Vera C. Rubin} Observatory’s Legacy Survey of Space and Time (Rubin-LSST\footnote{\url{https://rubinobservatory.org/}}; \citealt{Ivezic2019LSST}).

The remainder of this paper is organized as follows. In Section~\ref{sec:data}, we describe the JWST datasets used as the training basis for constructing \Roman-like galaxies. Section~\ref{sec:sim} explains the methodology for transforming JWST observations into synthetic \Roman\ images. In Section~\ref{sec:DDPM}, we present the diffusion model framework developed to generate realistic galaxy postage stamps at \Roman\ resolution. Section~\ref{sec:results} validates the fidelity of the generated sample, both through visual inspection and by comparing structural and photometric parameter distributions with those of the validation dataset. Finally, Section~\ref{sec:conclusion} summarizes our results and outlines their relevance for \Roman\ weak lensing analyses.

\section{Data: JWST Observations} \label{sec:data}
The dataset used in this study is based on observations from JWST, specifically from the JWST Advanced Deep Extragalactic Survey (JADES; \citealt{eisenstein2023}). JADES NIRCam covers the GOODS-S and GOODS-N extragalactic fields with ultra-deep and medium-depth imaging, providing high-resolution observations over a total area of 67 arcmin$^{2}$ in GOODS-S (27 arcmin$^{2}$ in the Deep program and 40 arcmin$^{2}$ in the Medium program) and 58 arcmin$^{2}$ in GOODS-N. The observations span the wavelength range $0.8$-$5.0,\mu$m across a suite of nine broad and medium NIRCam filters, reaching depths as faint as $m_{\rm AB} \approx 30.6$ (5$\sigma$) in the deepest fields at $2.7,\mu$m. The mosaics have a pixel scale of $0.0312^{\prime\prime}$/pix, with a PSF full width at half maximum (FWHM) of $\sim0.07$-$0.15^{\prime\prime}$ depending on filter.

For the purposes of this work, we restrict our analysis to NIRCam imaging in the F090W, F115W, F150W, and F200W filters in both GOODS-S and GOODS-N, selecting only sources with coverage in all four bands. The transmission curves of these JWST filters, together with the \Roman\ Y, J, and H filters, are shown in Fig.~\ref{fig:trans_curve}. These JWST filters overlap with the set of bands adopted for the \Roman\ HLS, namely Y (F106), J (F129), and H (F158). \Roman\ will employ these three near-IR bands in the medium-tier for WL analyses, as they enable multiple independent shape measurements, mitigate chromatic PSF biases when combined with optical data, and provide accurate photometric redshifts.

We adopt a similar methodology to that described in \citet{Scognamiglio_2025} to generate $198 \times 198$ pixel postage stamps centered on galaxies in the GOODS-S and GOODS-N fields. Galaxies are selected from the publicly available JWST NIRCam catalogs based on uniform criteria across the F090W, F115W, F150W, and F200W bands to ensure reliable photometry and morphology measurements. We exclude stars using the catalog star flag (\texttt{flag\_st} $\neq 1$). To retain sources with sufficient signal-to-noise, we require magnitudes brighter than $\texttt{hmag} < 27.0$, where the magnitude cut is applied specifically to the \texttt{F200W\_KRON} flux, converted to AB magnitude. This choice ensures selection in the deepest band while maintaining completeness across the other filters. We impose a minimum half-light radius of $r_{\rm{half}}> 0.1''$, where $r_{\rm{half}}$ corresponds to the catalog quantity \texttt{F200W\_RHALF} (given in arcseconds), ensuring galaxies are well resolved relative to the NIRCam PSF.
Finally, we require an axis ratio $b/a > 0.3$ to exclude extremely elongated sources, which are often spurious detections or artifacts near the noise limit.

Unlike \citet{Scognamiglio_2025}, we do not apply cuts in redshift; instead, we include galaxies across the full range to preserve the diversity of morphologies, sizes, and surface brightnesses, consistent with the broad source distribution expected for \textit{Roman} WL surveys.

This procedure yields 9,959 postage stamps in each of GOODS-S and GOODS-N, for a total of 19,888 cutouts. At this stage, projected neighboring sources within each cutout are intentionally kept and are removed only later during the \Roman-like interloper-masking procedure described below.

\section{Generating \Roman-like Galaxies}
 \label{sec:sim}
\begin{figure}
    \centering  \includegraphics[width=\columnwidth]{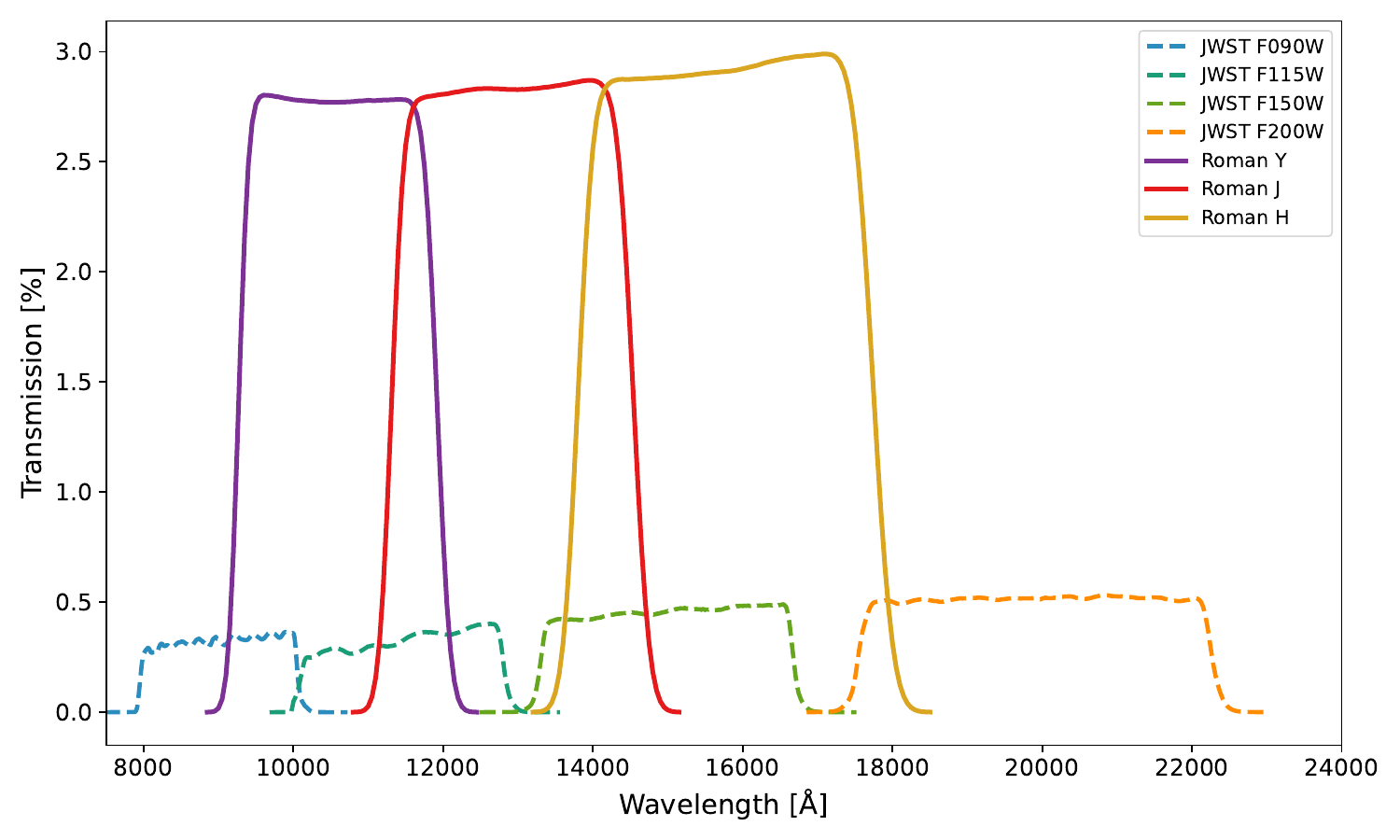}    
    \caption{Transmission curves of the \Roman\ WFI filters (Y, J, H) and JWST NIRCam filters (F090W, F115W, F150W, F200W) used to simulate \Roman-like galaxies based on JWST galaxy images.}   \label{fig:trans_curve}
\end{figure}

\Roman-like galaxies are generated from JWST/NIRCam cutouts in the F090W, F115W, F150W, and F200W filters, following a procedure similar to that described in \citet{Scognamiglio_2025}. The overlap between JWST and \Roman\ filters is illustrated in Fig.~\ref{fig:trans_curve}, which motivates the linear combination used to construct \Roman\ Y, J, and H images from the NIRCam bands.

For each stamp, a local background level is estimated from the outer border of the image and subtracted. To reduce high-frequency noise while preserving fine morphological structure, we apply a Wiener filter with a $5 \times 5$ kernel.

The NIRCam PSFs are taken from the STScI STPSF library \citep{Perrin2014}, while the \Roman\ PSFs are generated in \texttt{GalSim} \citep{Rowe2015} as monochromatic obscured Airy patterns at the central wavelength of the corresponding band. This corresponds to an intentionally simplified, spatially invariant PSF model that does not include wavelength-dependent effects across the bandpass. This approximation is adopted to isolate the performance of the galaxy-generation pipeline without introducing additional complexity from instrumental modeling. In particular, the current implementation does not account for chromatic PSF effects arising from the convolution of wavelength-dependent PSFs with galaxy spectral energy distributions (SEDs). In realistic \Roman\ observations, such chromaticity can introduce multiplicative shear biases at the level of $m \sim 10^{-3}$--$10^{-2}$, exceeding the \Roman\ WL requirement of $|m| \lesssim 3\times10^{-4}$ if uncorrected (e.g., \citealt{Berlfein_2025}). Because the same PSF model is applied consistently to the training, validation, and generated samples, the omission of chromatic effects does not impact the internal consistency of the validation presented in this work. However, it limits the realism of the simulations for applications that require accurate shear calibration, where PSF-SED coupling can modify the effective galaxy shapes. Incorporating wavelength-dependent and spatially varying PSFs, together with exposure-level effects such as dithering, will be an important extension of this work toward fully survey-realistic WL simulations.

The primary aperture diameter is set to 2.36 m, with a central obscuration factor of about 0.376 (corresponding to 14.1\% obscured area), consistent with the effective collecting area of the \Roman\ WFI. Each cutout is deconvolved from its native NIRCam PSF and reconvolved with the \Roman\ PSF, effectively translating the observed morphology into the \Roman\ observing regime.

Pixel scales are adjusted by converting the NIRCam detector resolution of 0.0312 arcsec/pixel to the \Roman\ WFI scale of 0.11 arcsec/pixel. As a result, a $198 \times 198$ NIRCam stamp becomes $56 \times 56$ pixels, corresponding to $6.16'' \times 6.16''$ on the sky. 

\Roman-like Y, J, and H images are constructed by linearly combining the PSF-matched NIRCam bands using throughput-based weights derived from the overlap of the filter transmission curves (Fig.~\ref{fig:trans_curve}). 
Because the intermediate NIRCam bands F162M and F182M are not included, their contributions are set to zero. Based on the transmission curves, this omission has negligible impact on the Y band, a sub-percent ($\sim$0.3\%) effect on J, and a more significant impact on H, where the effective throughput is reduced by $\sim$57\% relative to full NIRCam wavelength coverage. The same procedure is applied consistently to both training and validation datasets. To assess whether this bandpass mismatch introduces selection effects, we performed diagnostic tests comparing fluxes reconstructed using the available JWST bands to those obtained when including all available JWST bands (i.e., a more complete bandpass coverage). We find that the resulting flux differences follow a unimodal and narrowly peaked distribution, with no evidence for a distinct subpopulation of galaxies. This indicates that the bandpass gap primarily introduces a smooth, well-behaved flux bias affecting the normalization, rather than a significant incompleteness in the galaxy population.


Before the \Roman-like images are used for training or validation, we perform interloper masking to remove projected neighbours while preserving the statistical properties of the background. For each three-band Y-J-H stamp, we construct a detection image by summing the three bands and run Source Extractor (SEP; \citealt{Bertin1996, Barbary2016}) to obtain a segmentation map. The central galaxy is identified as the object whose segmentation label includes the geometric center of the cutout. To ensure that small-scale structures and close blends associated with the central galaxy are preserved, this central label is dilated slightly, and any neighboring labels touching it are retained. All other detected objects are classified as interlopers.

An interloper mask is then defined from these labels and dilated by a few pixels so that the full light distribution of each contaminant is removed. A second, more conservative mask---obtained by dilating the full segmentation map---is used to exclude sources when estimating the background. For each band, we estimate a local background level and a robust background noise standard deviation using 3-pixel–wide stripes along the outer edges of the image, selected to exhibit background-like statistics.
To reproduce the noise correlation present in the original stamp, we create a \texttt{GalSim  CorrelatedNoise} model using a background-only version of the image, in which all detected sources are replaced by the estimated global background and the mean is removed. A correlated-noise realization is drawn from this model, rescaled so that its standard deviation matches that measured in the stripe region, and shifted to ensure zero mean over the interloper pixels. Finally, the interloper regions in each band are replaced with this correlated noise added to the local background. This procedure eliminates neighboring contaminants while maintaining both the spatial noise correlation and the global noise amplitude of the \Roman-like stamp.

\begin{figure}
    \centering  \includegraphics[width=\columnwidth]{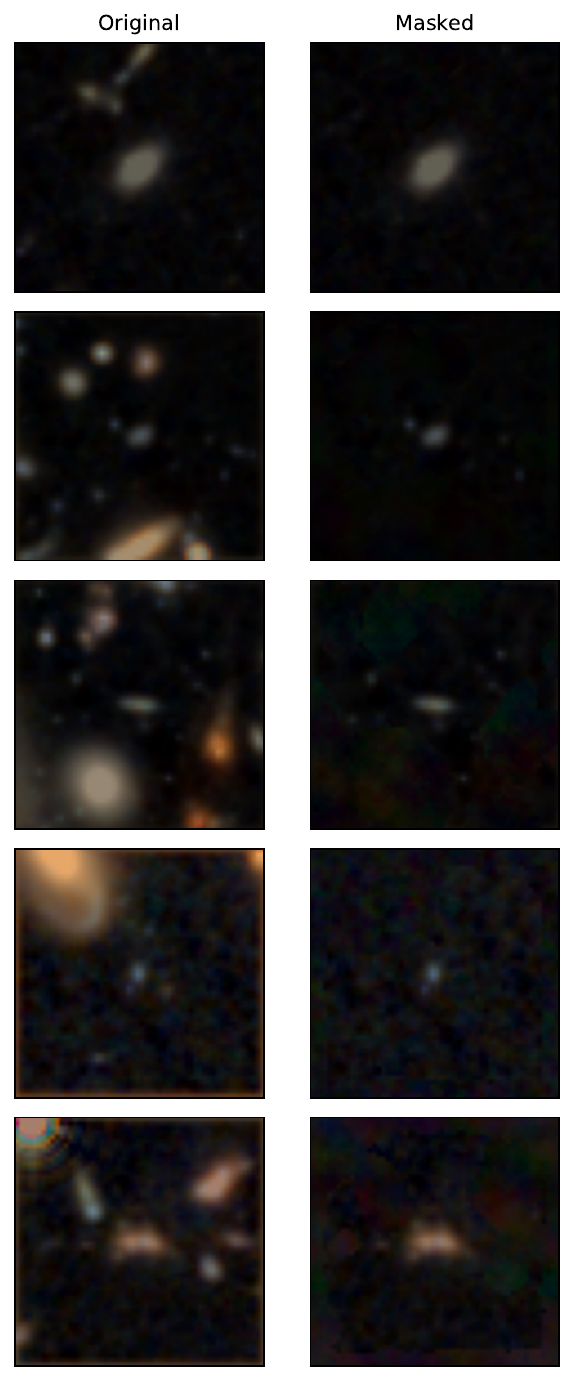}    
    \caption{Examples of the interloper-masking procedure. Left: original RGB composites (R = H, G = J, B = Y). Right: corresponding masked stamps, where projected neighbors are removed and replaced with locally matched correlated noise. The central galaxy morphology is preserved.}   \label{fig:orig_mask}
\end{figure}

We visually inspect the masking results by constructing RGB composites (R = H, G = J, B = Y), as displayed in Fig.~\ref{fig:orig_mask}. In the original images, projected neighbors and bright companions are clearly visible, while in the masked versions these contaminants are removed and replaced by statistically consistent correlated noise, leaving the central galaxy morphology untouched.

The resulting \Roman-like dataset consists of 19,888 masked Y, J, and H cutouts. However, the final datasets used for training and validation are obtained after applying additional measurement-based quality cuts on these \Roman-like stamps. For each stamp and for each band (Y, J, H), we measure central-object properties using \texttt{Photutils} segmentation photometry. We require that the central object is successfully detected and measured in all three bands, forming a strict three-band matched sample $(Y \cap J \cap H)$.

We then apply a resolved-size cut using a moment-based size proxy computed from the \texttt{Photutils} second-moment ``sigma'' sizes. Specifically, \texttt{Photutils} provides the \texttt{semimajor\_sigma} and \texttt{semiminor\_sigma} (in pixels), and we define a circularized size proxy $R_{\rm{m}} = (\texttt{semimajor\_sigma} + \texttt{semiminor\_sigma)}/2$. We require $R_{\rm{m}} \geq 2.0$ pixels in each band, and we keep only stamps passing this cut in all three bands.

After applying these cuts, we split the resulting datasets into training and validation datasets. The final samples contain 10,870 galaxies for training and 2,734 galaxies for validation. The validation stamps are drawn from the same parent distribution but do not overlap with the training sample, providing an independent benchmark for assessing the fidelity of the generated \Roman-like galaxies (see Sect.~\ref{sec:results}).

\section{Generative Model for \Roman-like Simulations}
\label{sec:DDPM}
\begin{figure*}[t]
    \centering
    \includegraphics[width=\textwidth]{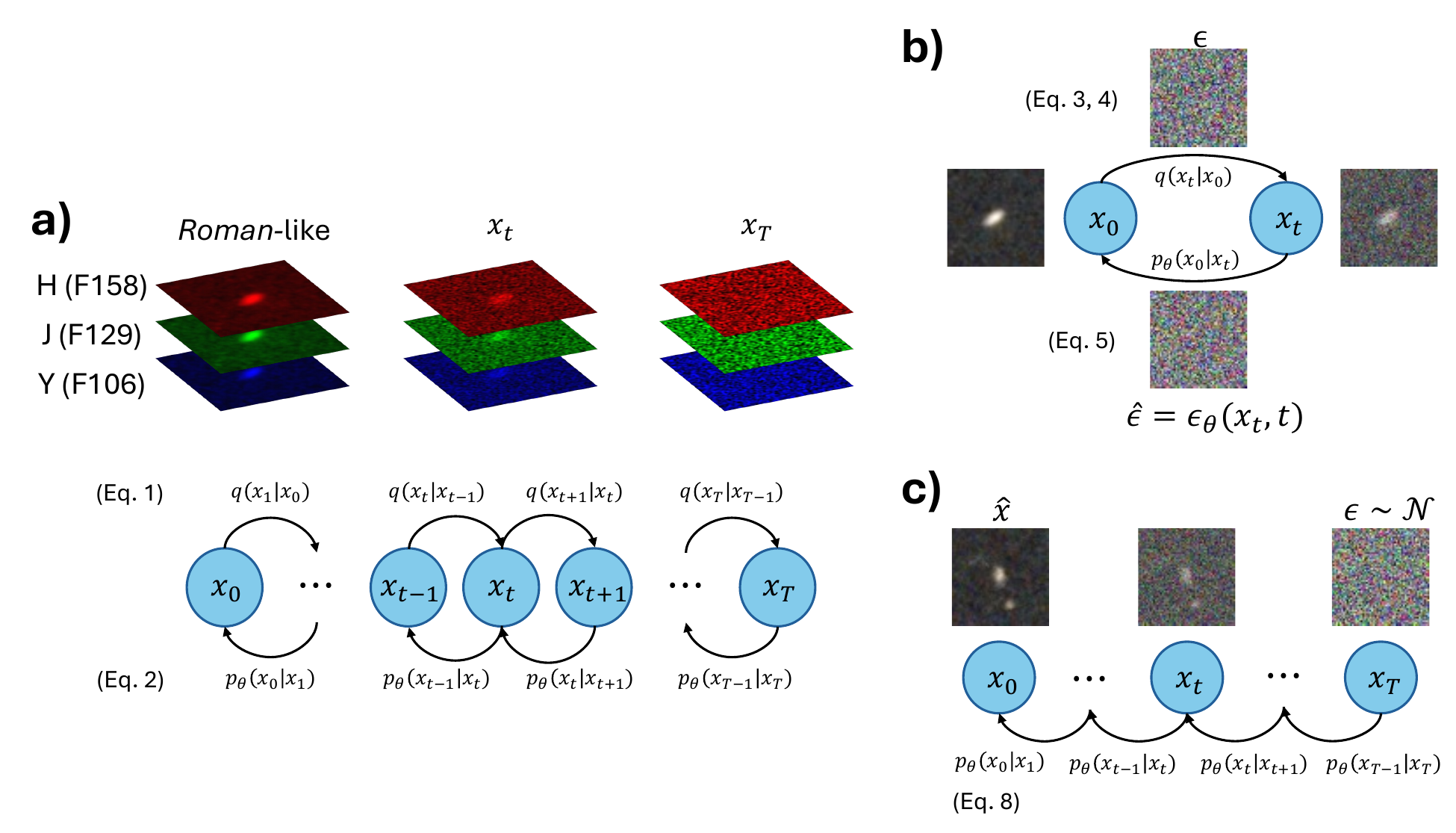}
    \caption{a) Illustration of the Denoising Diffusion Probabilistic Model (DDPM) architecture. 
    b) Training diagram for the parametrized model $\epsilon_\theta$. 
    c) Sampling procedure.}
    \label{fig:ddpm}
\end{figure*}

Denoising Diffusion Probabilistic Models (DDPMs or ``diffusion models''; \citealt{ho2020DDPM, sohl-dickstein15}) are a class of deep generative models that has demonstrated state-of-the-art performance in image generation. Compared to prior methods such as Generative Adversarial Networks (GANs; \citealt{goodfellow2014GAN}), diffusion models generate higher-quality images that better resemble the distribution of training data \citep{dhariwal2021diffusion}. While our work focuses on unconditional image generation, diffusion models have also demonstrated superior performance in conditional image generation \citep{ho2022classifier}, which have been successfully applied to other cosmological applications \citep{mudur2023cosmological, riveros2025conditional}. 

The diffusion model, illustrated in Fig.~\ref{fig:ddpm}, is defined as a Markov chain of the forward diffusion process and the reverse denoising process. The forward process is a fixed Markov chain that incrementally adds Gaussian noise to a given \Roman-like image $\mathbf{x}_0$ sampled from the training data distribution $q(\mathbf{x})$ over $T$ steps:
\begin{equation}
    q(\textbf{x}_{t}|\textbf{x}_{\textbf{t}-1})=\mathcal{N}(\textbf{x}_{t};\mathbf{\mu}_{t}=\sqrt{1-\beta_{t}}\textbf{x}_{t-1},\mathbf{\Sigma}_{t}=\beta_{t}\textbf{I}),
\end{equation}
where $\bm{\mu}_{t}$ and $\mathbf{\Sigma}_{t}$ are the mean and the variance of the distribution, respectively, $\beta_{t}$ is the noise variance, equal for each dimension of the multi-dimensional space, and $\mathbf{I}$ is the identity matrix. Note that $\mathbf{x}_{T}$ at the end of this chain is pure noise.

The reverse process aims to recover the original data by incrementally denoising starting from $\mathbf{x}_{T}$. The true reverse distribution $q(\mathbf{x}_{t-1} \vert \mathbf{x}_t) $ cannot be computed because it depends on the entire dataset's distribution $ q(\mathbf{x}_t) $, which is unknown. Instead, it is approximated with a parameterized model $ p_\theta(\mathbf{x}_{t-1} \vert \mathbf{x}_t) $, where $ \theta $ represents neural network parameters learned to estimate the ideal (but intractable) reverse process. This network predicts denoising steps using only the current noisy input $ \mathbf{x}_t $. Since $q(\mathbf{x}_t | \mathbf{x}_{t-1})$ is Gaussian, for small $\beta_{t}$, $p_{\theta}$ can also be considered Gaussian and parameterize the mean and variance as follows:
\begin{equation}
    p_{\theta}(\mathbf{x}_{t-1}|\mathbf{x}_{t})=\mathcal{N}(x_{t-1};\bm{\mu}_{\theta}(\mathbf{x}_{t},t),\mathbf{\Sigma}_{\theta}(\mathbf{x}_{t},t)).
\end{equation}

Training optimizes the variational bound on the negative log likelihood. In implementation, we sample a time step $t$ and add Gaussian noise to a batch of images $\mathbf{x_0}$ to produce $\mathbf{x_t}$:
\begin{align}
    \bar\alpha_t &= \Pi_{i=1}^t 1 - \beta_t \\
    x_t &= \sqrt{\bar\alpha_t}x_0 + \sqrt{1-\bar\alpha_t}\epsilon, \epsilon \sim \mathcal{N}(0,\mathbf{I}).
\end{align}

Note that the Gaussian forward process allows us to sample $x_t$ from $x_0$ directly without traversing the intermediate steps of the Markov chain via the additive property of Gaussians. A neural network $\epsilon_\theta$ now predicts the noise component, and the loss (mean squared error in this case) between the true and predicted noise is backpropagated with a gradient descent step to update $\theta$:
\begin{align}
    \hat\epsilon = \epsilon_\theta(x_t, t) \\
    \mathcal{L} = \| \epsilon - \hat\epsilon \|^2 \\
    \theta \leftarrow \eta\nabla_\theta\mathcal{L},
\end{align}
where $\eta$ is the learning rate. 

Sampling from a trained diffusion model, on the other hand, requires that we start with pure Gaussian noise $x_T$ and step backwards through the entire Markov chain. This can be computationally expensive, especially for large values of $T$. Instead, we utilize DPM-Solver++, a fast high-order solver for diffusion ordinary differential equations (ODEs, \citealt{lu2022dpm, Lu_2025}). Briefly, DPM-Solver++ reformulates the reverse diffusion process as a deterministic probability flow ODE determined by the trained noise predictor $\epsilon_\theta$:
\begin{equation}
    \frac{\dd\mathbf{x}_t}{\dd t} = f(t)\mathbf{x}_t + \frac{g^2(t)}{2\sigma_t}\epsilon_\theta(\mathbf{x}_t, t),
\end{equation}
where $f(t)$ and $g(t)$ represent the time-dependent drift and diffusion coefficients, respectively, derived from the noise schedule $\beta_t$. By analytically integrating the linear portion of the equation and approximating the nonlinear neural network prediction via high-order polynomial fitting, DPM-Solver++ minimizes discretization error. We further adopt the noise schedule and time-step spacing proposed by \citet{karras2022elucidating}, which optimizes the curvature of the ODE trajectory for faster convergence. Using the 3rd order solver with this configuration, we are able to generate galaxy stamps in as few as 25 steps without any loss in generated image quality.

Our model was implemented with the Hugging Face Diffusers library \citep{von-platen-etal-2022-diffusers}, which abstracts most of these concepts into single-line API calls. The parametrized model $\epsilon_\theta$ is a U-Net \citep{ronneberger2015u} with an input dimension $3 \times 56 \times 56$ and three two-layer convolutional blocks of 128, 256, and 512 channels. We found that increasing model complexity with additional channels or attention blocks degraded the alignment between the generated and training distributions of galaxy metrics. We set diffusion steps $T=500$ with a linear $\beta$ noise schedule; decreasing or increasing the number of steps resulted in worse image fidelity and insignificant improvements at greater compute cost, respectively. The model is trained with the AdamW optimizer \citep{Loshchilov2017DecoupledWD} on Mean Squared Error loss with an initial learning rate of $10^{-4}$ following a cosine schedule for 100 epochs.

To improve model training stability and model expressiveness, we normalize the data to the range $[-1, 1]$ with the following transformation:
\begin{align}
    \mathbf{x} &= 4\,\sigma \cdot \mathrm{asinh}\!\left(\frac{\mathbf{x}}{3\sigma}\right) \\
    \sigma &= 0.02297293 \nonumber
\end{align}
During sampling, the inverse of this transformation is applied to reproduce the original magnitudes. In addition to gradient stability during back-propagation, this transformation is motivated by the fact that the noise at $\mathbf{x}_T$ is sampled from the standard normal distribution.

The model takes 1 hour to train for 92,298 steps with a batch size of 16 on a single NVIDIA H100 GPU. For inference, 10,000 galaxies can be generated in 159 seconds, or 0.0159 seconds per stamp, on the same hardware. Sampling with DPM++ Karras and $T=25$ reduces our inference runtime by an order of magnitude compared to standard sampling with $T=500$.

The complete codebase for dataset preparation, model training, and galaxy generation is publicly available on GitHub\footnote{\url{https://github.com/diana-sco/roman-galaxy-ddpm}}. The train, test, and generated datasets, together with trained model weights and a packaged version of the software, are archived on Zenodo \citep{roman_ddpm_zenodo}.

\section{Validation of Generated Data} \label{sec:results}
\begin{figure*}[h!]
    \centering
    \textbf{DDPM-generated \Roman-like galaxies (Gen)}\\
    \includegraphics[width=0.72\textwidth]{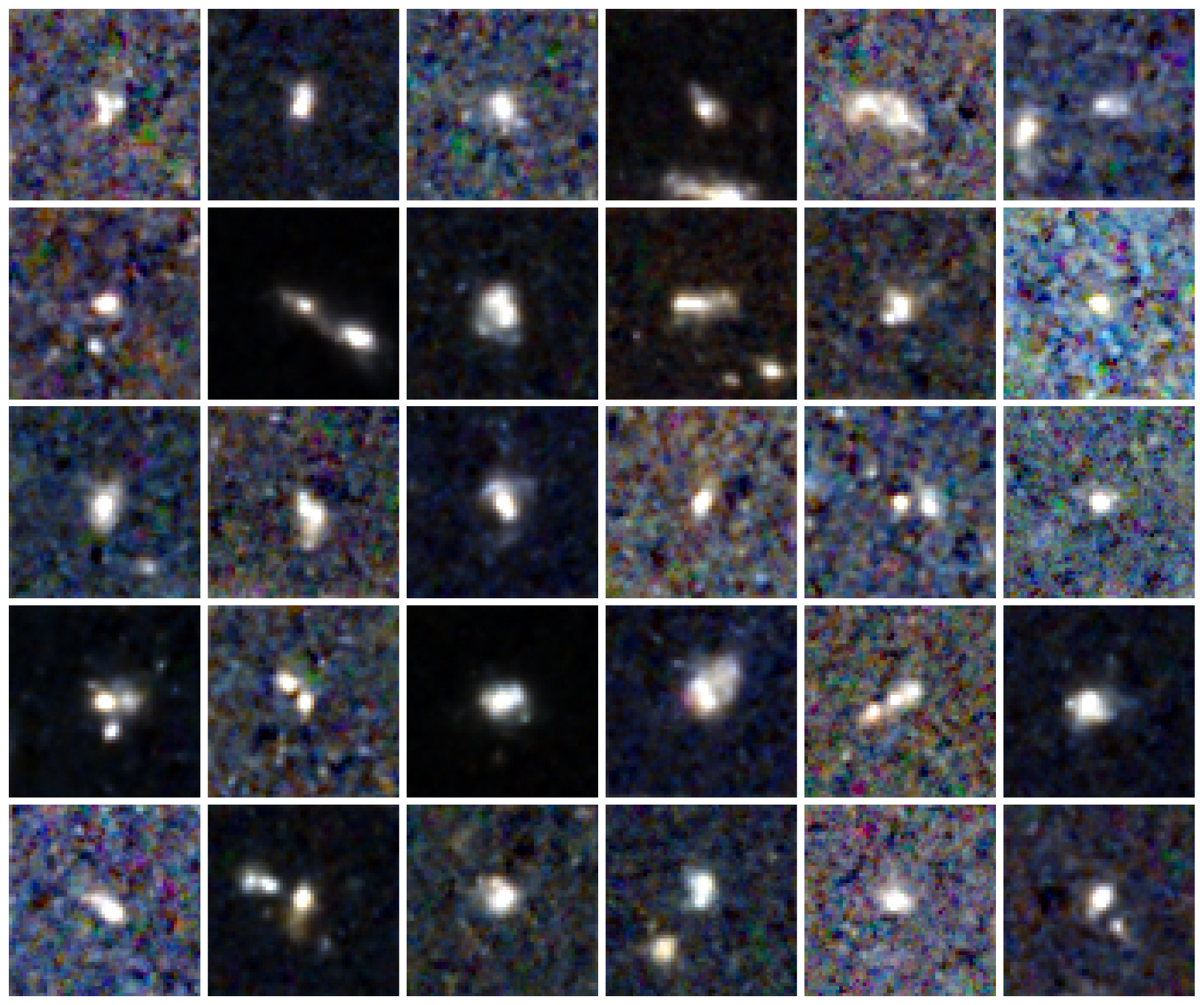}\\
    \vspace{0.4cm}
    \textbf{\Roman-like galaxies (Val)}\\
    \includegraphics[width=0.72\textwidth]{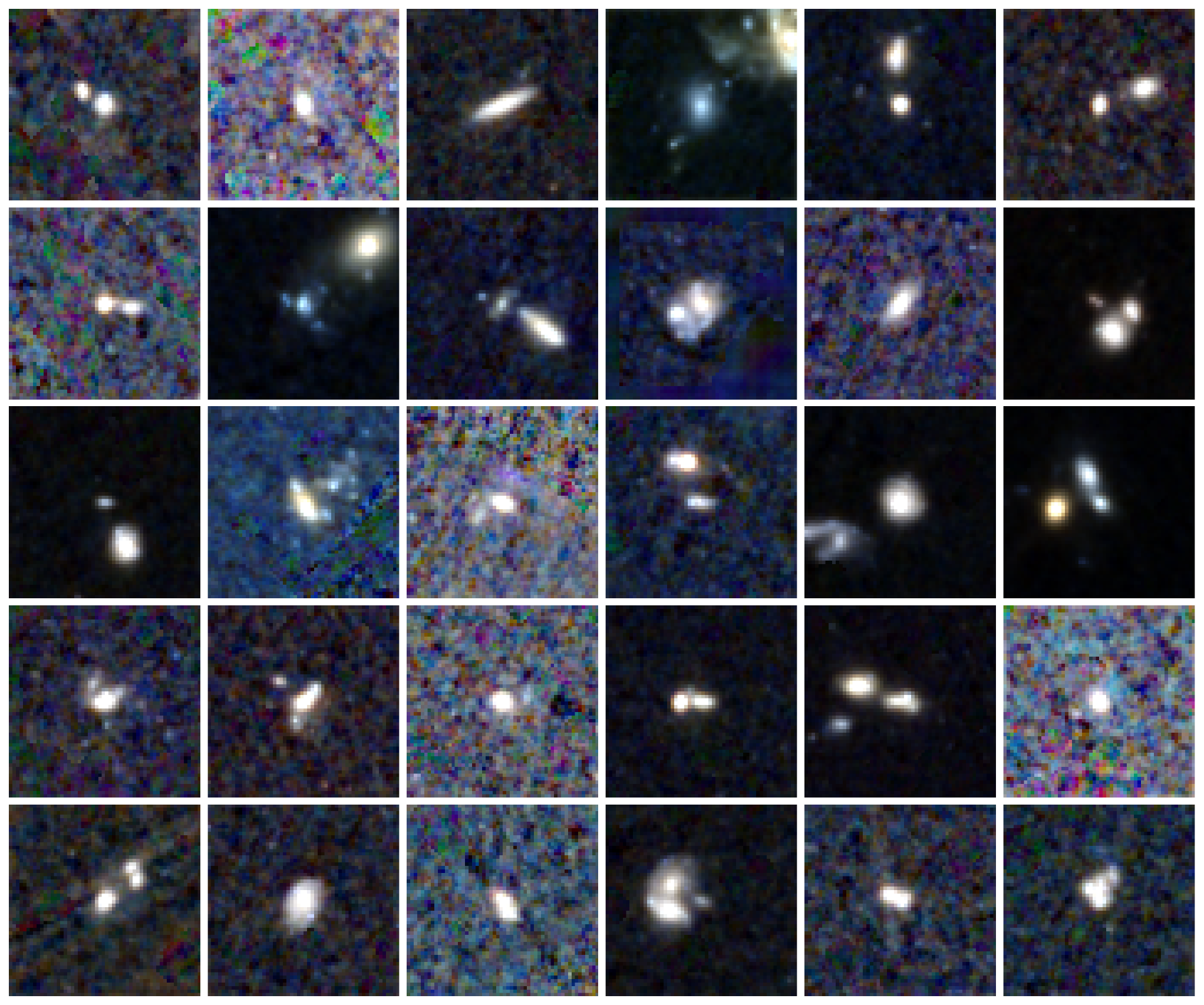}
    \caption{Comparison between the DDPM-generated \Roman-like galaxies (Gen; top) and validation galaxies (Val; bottom). The generated images are produced using the DDPM model trained on \Roman-like galaxy data, exhibiting realistic morphological features consistent with the training set. Each postage stamp is 56 $\times$ 56 pixels, corresponding to $6.16^{\prime\prime} \times 6.16^{\prime\prime}$ at the \Roman\ pixel scale.}
    \label{fig:gal_comparison}
\end{figure*}

\begin{figure*}[t!]
    \centering
    \textbf{Y band}\\
    \includegraphics[width=\textwidth]{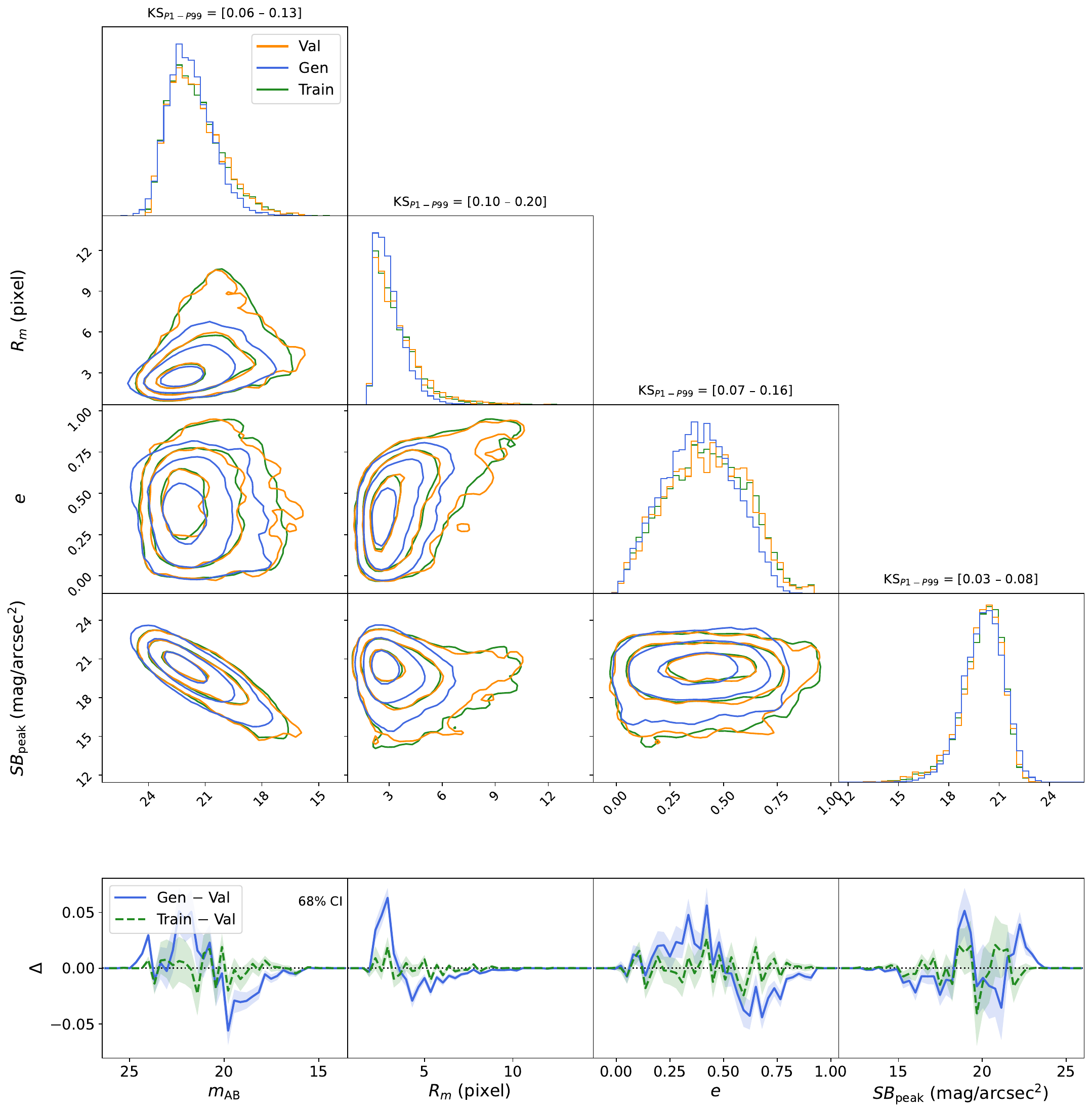}
    \vspace{0.5cm}
    \caption{
    Comparison of galaxy property distributions between the validation (Val), generated (Gen), and training (Train) samples in the Y band. The upper panels display the joint one- and two-dimensional distributions of Kron AB magnitude $m_{\mathrm{AB}}$, moment-based radius $R_m$ (pixel), ellipticity $e$, and peak surface brightness $SB_{\mathrm{peak}}$ (mag arcsec$^{-2}$). The diagonal panels show normalized one-dimensional histograms, while the off-diagonal panels present two-dimensional density contours derived via kernel density estimation, illustrating the covariance between parameters. The diagonal panels also report the bootstrap range of the Kolmogorov--Smirnov (KS) statistic [P$_1$--P$_{99}$] computed between samples. The lower panels show the residuals of the one-dimensional distributions, defined as $\Delta \equiv p_1 - p_2$, where $p$ denotes the normalized histogram density in each bin. Solid blue lines indicate Gen--Val differences, and dashed green lines indicate Train--Val differences. Shaded regions represent the 68\% confidence intervals estimated via bootstrap resampling.
    }
    \label{fig:hist_train_val_Y}
\end{figure*}

\begin{figure*}[t!]
    \centering
    \textbf{J band}\\
    \includegraphics[width=\textwidth]{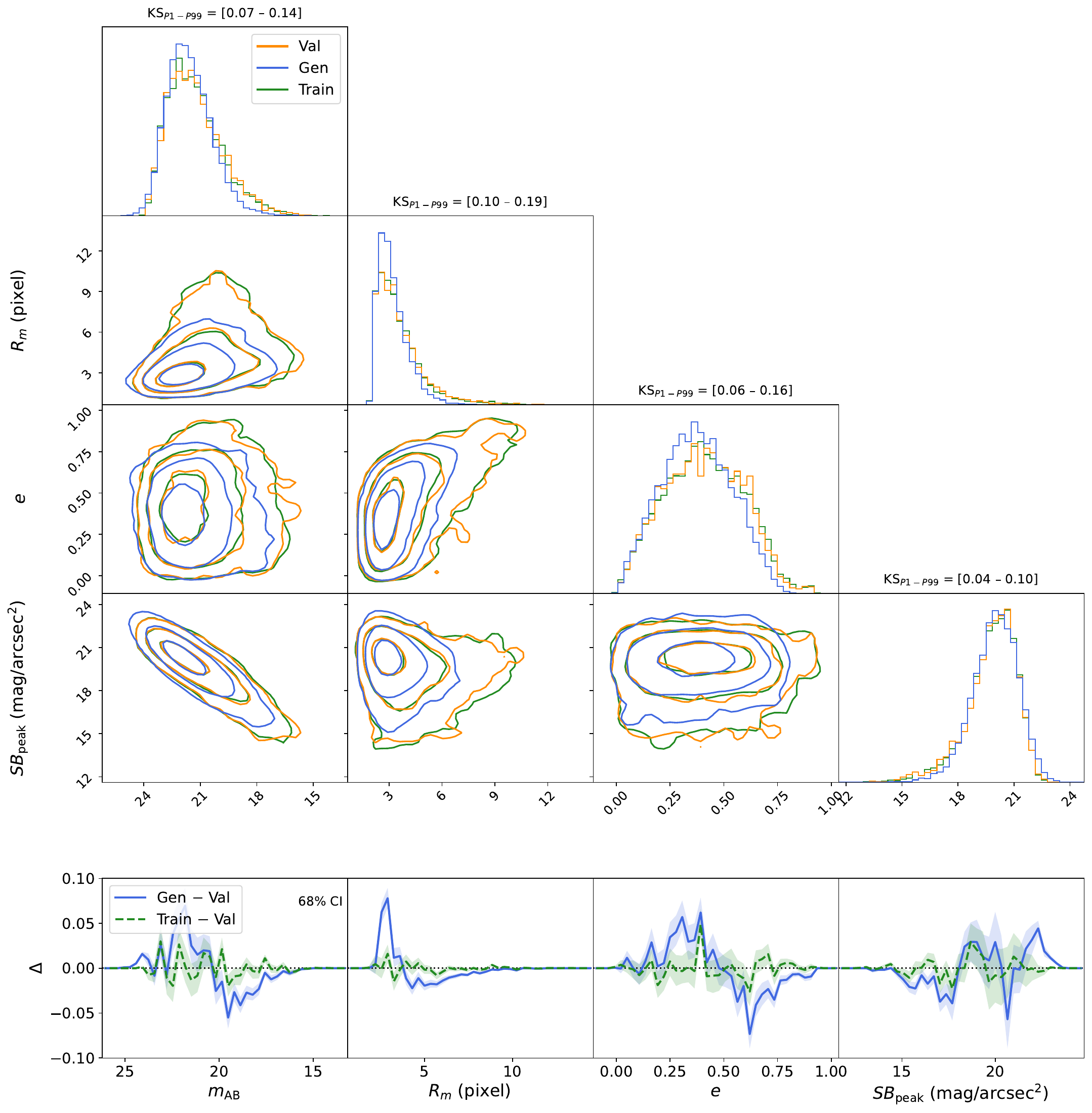}
    \vspace{0.1cm}
    \caption{Same as Fig.~\ref{fig:hist_train_val_Y}, but for the J band.}
    \label{fig:hist_train_val_J}
\end{figure*}

\begin{figure*}[t!]
    \centering
    \textbf{H band}\\
    \includegraphics[width=\textwidth]{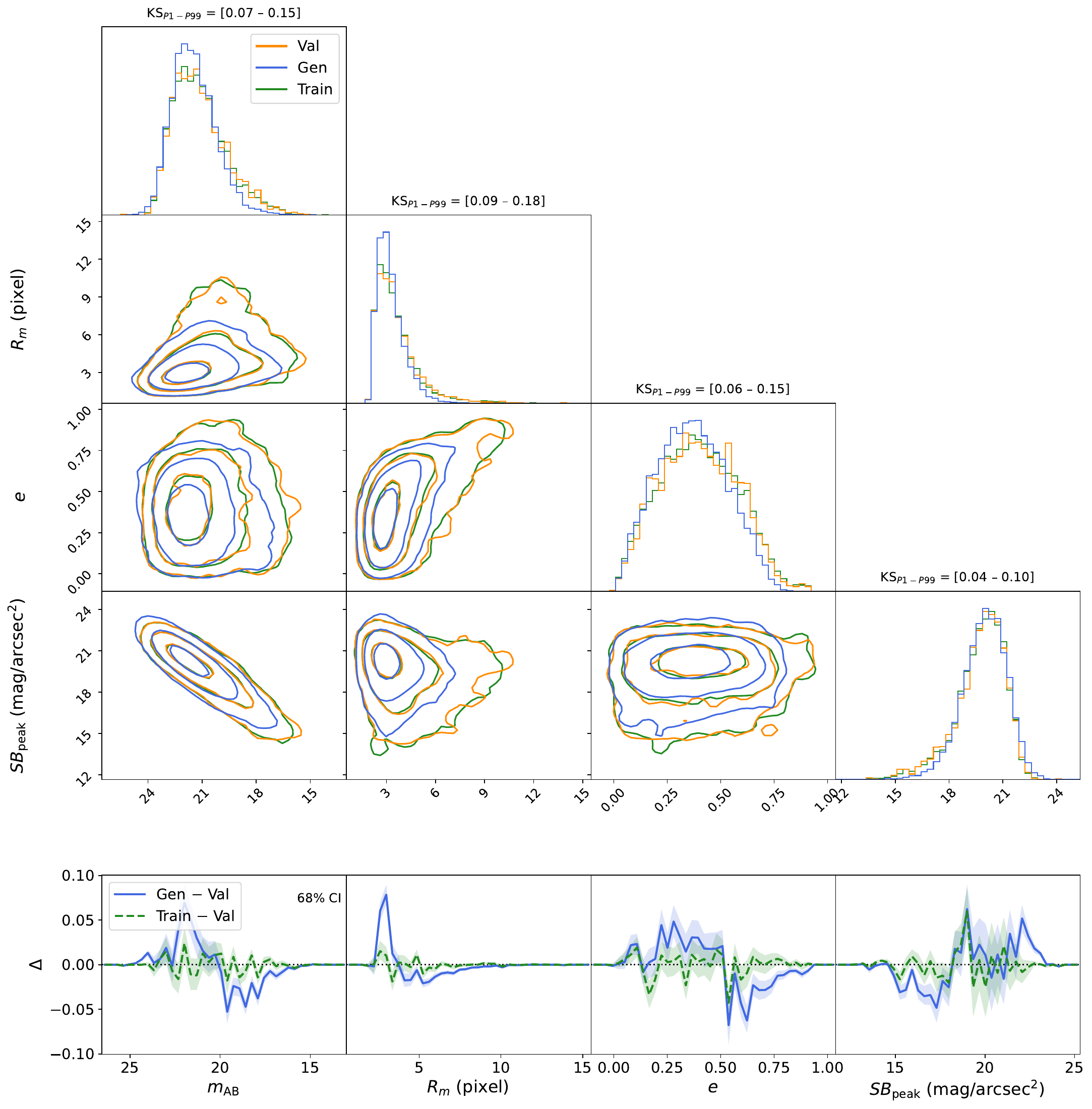}
    \vspace{0.5cm}
    \caption{Same as Fig.~\ref{fig:hist_train_val_Y}, but for the H band.}
    \label{fig:hist_train_val_H}
\end{figure*}

    
  
    
    
    

\begin{figure*}[t]
    \centering
\includegraphics[width=\textwidth]{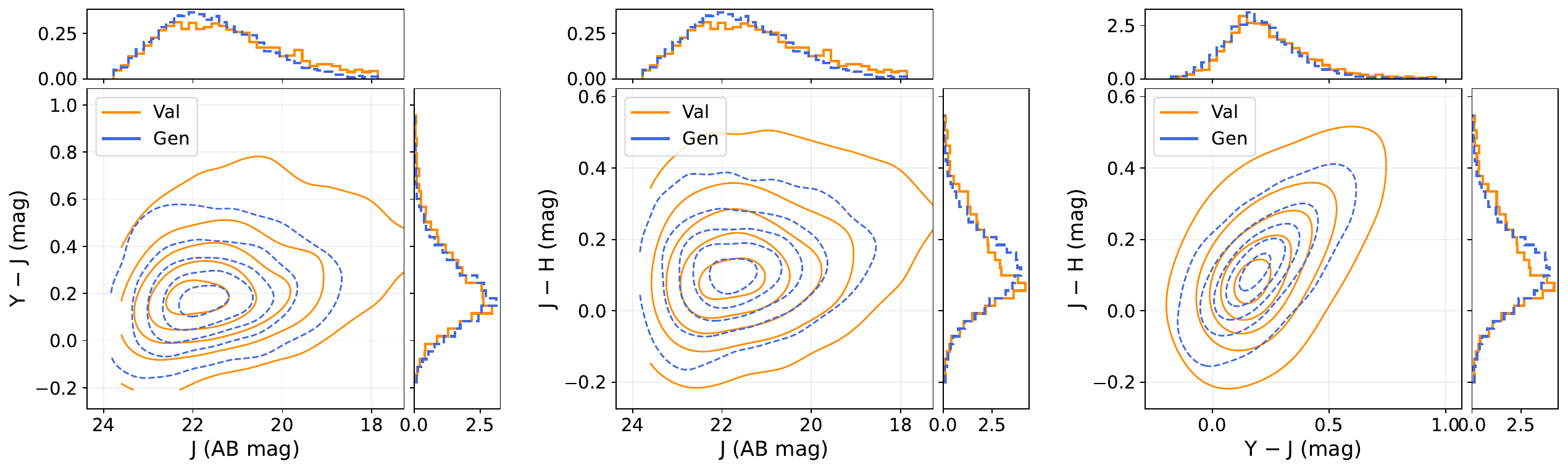}
    \caption{Color-magnitude and color-color distributions for the validation (Val) and generated (Gen) samples. The panels show Y--J and J--H as a function of J--band magnitude (left and middle) and the Y--J versus J--H relation (right). Density contours illustrate the two-dimensional distributions, with normalized one-dimensional marginals shown along the top and right of each panel.}
    \label{fig:color-color}
\end{figure*}

Validating the realism of the diffusion-generated galaxies is essential to ensure that they can be safely used for \Roman\ weak lensing analyses and related applications. We therefore compare the DDPM-generated images (Gen) to an independent validation (Val) sample of \Roman-like stamps derived from JWST/NIRCam cutouts (Sect.~\ref{sec:sim}).
The validation follows a single photometric pipeline applied identically to Train, Val, and Gen, and combines (i) visual inspections of three-band RGB postage stamps, (ii) multivariate comparisons of structural and photometric measurements, and (iii) distribution-level statistical tests.

\subsection{Measurement Pipeline and Sample Definition}
All measurements are performed independently on each postage stamp in the Y, J, and H bands. For each stamp, we first estimate and subtract a two-dimensional background using \texttt{photutils} \texttt{Background2D} with a median background estimator. The stamp is split into $10 \times 10$ pixel tiles, and the resulting coarse background grid is smoothed with a $3 \times 3$ median filter to suppress tile-scale fluctuations before interpolation to full resolution. We subtract this background model prior to source detection.

Source detection is performed on the background-subtracted image after convolution with a circular 2D Gaussian kernel ($\sigma = 3$ pixels; kernel size $= 5$ pixels). We identify sources 
and request at least 10 connected pixels. Among all detected segmentation labels, we define the \emph{central galaxy} as the object whose segmentation-region centroid is closest to the geometric center of the stamp. All subsequent measurements are derived exclusively for this central object.

For each central galaxy we measure: (1) Kron flux $F$ and its uncertainty (\texttt{kron\_flux}, \texttt{kron\_fluxerr}, \citealt{Kron1980}) evaluated on the background-subtracted image; (2) the moment-based radius $R_{\rm m}$ as defined in Sect.~\ref{sec:sim}, computed from the second-moment dispersions (via \texttt{semimajor\_sigma} and \texttt{semiminor\_sigma}); (3) ellipticity defined as $e = 1 - b/a$; and (4) the peak surface brightness 
$SB_{\rm peak}$, defined as the maximum background-subtracted pixel value within the segmentation region of the central object. Because the stamps are stored in surface-brightness units (MJy/sr), we convert Kron flux to flux density in Jy and we compute AB magnitudes $m_{\rm AB}$ from this flux density. Throughout the validation figures and statistical comparisons, we use $m_{\rm AB}$
(derived from the Kron flux) as the photometric variable rather than $F$
itself. Peak surface brightness values are converted from MJy/sr to AB mag/arcsec$^{2}$ using the same surface-brightness conversion adopted throughout the pipeline.

To ensure a consistent comparison across datasets and plots, we apply identical quality selections in each band and then enforce a three-band match within each dataset (Gen, Val, and Train). First, we require the measured centroid of the central galaxy to lie within a radius of $r_{\rm cut}$ pixels from the stamp center (here $r_{\rm cut} = 20$), mitigating failures where the closest-to-center label does not correspond to the intended object or where the main object lies near the stamp boundary. Second, we require a minimum size $R_{\rm m} \geq 2.0$ pixels. Third, we impose a minimum Kron signal-to-noise ratio $SNR \geq 3$, where $SNR = F/\sigma_{F}$ is computed from \texttt{kron\_flux} and \texttt{kron\_fluxerr}. Finally, we construct a three-band-matched catalog within each split by retaining only objects that pass the above cuts in Y, J, and H simultaneously
The DDPM model is used to generate an initial sample of 10,000 \Roman-like galaxy stamps. After applying the measurement pipeline and quality cuts described above, the final generated sample used for validation contains 7,538 galaxies. For comparison, the validation sample contains 2,734 galaxies and the training sample contains 10,870 galaxies after applying the same three-band consistency requirements.
The selection criteria applied here are closely aligned with those used to construct the \Roman-like dataset described in Sect.~\ref{sec:sim}, with the addition of centroid and signal-to-noise cuts to ensure robust measurements. The reduction in the generated sample size is primarily driven by detection failures and low signal-to-noise objects, and the requirement of consistent detections across all three bands.
After three-band synchronization, we verify that all retained objects have finite values for the plotted quantities ($m_{\rm AB}$, $R_{\rm m}$, $e$, $SB_{\rm peak}$) in all three bands.

\subsection{Visual and Statistical Validation}
We assess qualitative realism using RGB composites constructed from the three-band stamps. Fig.~\ref{fig:gal_comparison} shows randomly selected RGB postage stamps from the DDPM-generated (Gen; upper panel) and validation (Val; bottom panel) samples. The two sets display comparable levels of morphological diversity, including compact and extended systems as well as blended configurations. The visual similarity indicates that the diffusion model reproduces the overall appearance of the \Roman-like galaxy stamps at the stamp level.

We then compare the joint and marginal distributions of the measured parameters using combined one- and two-dimensional distribution panels with residual diagnostics. For each band, Figs.~\ref{fig:hist_train_val_Y}, \ref{fig:hist_train_val_J}, and \ref{fig:hist_train_val_H} show the two-dimensional density contours and one-dimensional marginalized distributions (upper panels), together with the corresponding one-dimensional residuals relative to the validation sample (lower panels), for $m_{\rm AB}$ (derived from the Kron flux), $R_{\rm m}$ (pixels), ellipticity 
$e$, and $SB_{\rm peak}$ (AB mag/arcsec$^{2}$), including the Val, Gen, and Train samples. Across all bands, Gen and Val samples show strong overall agreement in their marginal distributions. The $m_{\rm AB}$ histograms overlap closely, with well-aligned peaks and only small deviations in the faint tail. The moment-based radius $R_{\rm m}$ exhibits a mild systematic trend, with Gen slightly more concentrated toward smaller radii, particularly in Y and H. The ellipticity distribution is well reproduced in shape, with only minor differences at high $e$ ($\gtrsim 0.7$). The $SB_{\rm peak}$ distribution exhibits excellent agreement in all bands and represents one of the strongest agreements between the two samples. The Train sample follows the same overall trends as the Val sample, with comparable distributions and slightly larger deviations in low-density regions, providing a reference for the level of variation expected from finite sampling of the parent catalog.

The two-dimensional projections further demonstrate that the model preserves the underlying covariance structure of the data. In particular, the strong anti-correlation between $m_{\rm AB}$ and $\mathrm{SB}_{\rm peak}$ is recovered with nearly overlapping contours. Differences are modest and concentrated in the wings: Val typically exhibits slightly broader extensions toward larger radii, while Gen contours are marginally tighter, consistent with the mild compression observed in $R_{\rm m}$.

\begin{table*}[t]
\centering
\caption{Summary statistics for the validation (Val) and generated (Gen) samples in each band.
For each parameter we report the mean and standard deviation, the two-sample Kolmogorov--Smirnov statistic
\(D^{\rm KS}\), and the bootstrap range of \(D^{\rm KS}\) from the 1st to 99th percentiles, \(D^{\rm bootstrap\,KS}_{[1\text{--}99]}\).
Flux rows are excluded; the AB magnitude \(m_{\rm AB}\) is included.}
\label{tab:val_gen_stats}
\begin{tabular}{lcccc}
\toprule
\multicolumn{1}{c}{Parameter} &
\multicolumn{1}{c}{$\mu_{\rm Val}\pm\sigma_{\rm Val}$} &
\multicolumn{1}{c}{$\mu_{\rm Gen}\pm\sigma_{\rm Gen}$} &
\multicolumn{1}{c}{$D^{\rm KS}$} &
\multicolumn{1}{c}{$D^{\rm bootstrap\,KS}_{[1\text{--}99]}$} \\
\midrule
\multicolumn{5}{c}{\textrm{Y band}}\\
\midrule
$m_{\rm AB}$ & $21.51 \pm 1.38$ & $21.78 \pm 1.16$ & 0.0840 & [0.0580--0.1350] \\
$R_{\rm m}$ [pixel] & $3.46 \pm 1.30$ & $3.07 \pm 0.83$ & 0.1358 & [0.1010--0.1890] \\
$e$ & $0.421 \pm 0.182$ & $0.387 \pm 0.162$ & 0.0992 & [0.0690--0.1560] \\
$SB_{\rm peak}$ [mag arcsec$^{-2}$] & $19.94 \pm 1.31$ & $20.06 \pm 1.21$ & 0.0330 & [0.0280--0.0780] \\
\midrule
\multicolumn{5}{c}{\textrm{J band}}\\
\midrule
$m_{\rm AB}$ & $21.26 \pm 1.43$ & $21.57 \pm 1.18$ & 0.0946 & [0.0660--0.1480] \\
$R_{\rm m}$ [pixel] & $3.59 \pm 1.30$ & $3.21 \pm 0.83$ & 0.1284 & [0.0950--0.1840] \\
$e$ & $0.409 \pm 0.180$ & $0.375 \pm 0.160$ & 0.0976 & [0.0660--0.1500] \\
$SB_{\rm peak}$ [mag arcsec$^{-2}$] & $19.79 \pm 1.37$ & $19.99 \pm 1.23$ & 0.0482 & [0.0340--0.0920] \\
\midrule
\multicolumn{5}{c}{\textrm{H band}}\\
\midrule
$m_{\rm AB}$ & $21.12 \pm 1.47$ & $21.44 \pm 1.21$ & 0.0978 & [0.0670--0.1490] \\
$R_{\rm m}$ [pixel] & $3.62 \pm 1.29$ & $3.27 \pm 0.84$ & 0.1196 & [0.0830--0.1730] \\
$e$ & $0.399 \pm 0.178$ & $0.367 \pm 0.161$ & 0.0933 & [0.0560--0.1420] \\
$SB_{\rm peak}$ [mag arcsec$^{-2}$] & $19.74 \pm 1.42$ & $19.97 \pm 1.25$ & 0.0558 & [0.0400--0.1000] \\
\bottomrule
\end{tabular}
\end{table*}

To quantify distribution-level consistency, we compute the two-sample Kolmogorov–Smirnov (KS) statistic $D_{\rm KS}$ for each parameter and band (Table~\ref{tab:val_gen_stats}) for the Gen and Val samples. Because 
$D_{\rm KS}$ fluctuates with finite sampling, we estimate an expected variability range via bootstrap resampling: for each parameter we draw repeated bootstrap-resampled subsets from Val and Gen and compute $D_{\rm KS}$ for each realization. We report the 1st–99th percentile interval of the bootstrap $D_{\rm KS}$ distribution as a reference range for sampling variability. Table~\ref{tab:val_gen_stats} lists, for each band and parameter, the mean and standard deviation in Val and Gen, the measured $D_{\rm KS}$, and the bootstrap percentile interval $D_{[1-99]}^{\rm bootrstap~KS}$. Consistent with the contour plots, $SB_{\rm peak}$ yields the smallest $D_{\rm KS}$ values, ellipticity shows similarly good agreement, and the largest 
$D_{\rm KS}$ values occur for size, reflecting the mild tendency toward more compact generated objects. Overall, the measured $D_{\rm KS}$ values fall within the bootstrap ranges, indicating controlled differences.

To further quantify localized deviations, we examine the residuals of the one-dimensional distributions shown in the lower panels of Figs.~\ref{fig:hist_train_val_Y}, \ref{fig:hist_train_val_J}, and \ref{fig:hist_train_val_H}. The residuals are defined with respect to the validation sample as $\Delta_{\rm Gen} \equiv p_{\rm Gen} - p_{\rm Val}$ and $\Delta_{\rm Train} \equiv p_{\rm Train} - p_{\rm Val}$, where $p$ denotes the binned, normalized histogram density for each variable.
Confidence bands for each residual curve are estimated via bootstrap resampling and reported as 68\% confidence intervals. These residual diagnostics highlight where deviations are localized in parameter space and show that Gen–Val residuals are typically comparable to, or track, Train–Val residuals. This behaviour is expected when Train and Val are random subsets of the same parent catalog: by chance, the training split may under-sample some regions that are present in the validation set, especially in low-occupancy tails. Such finite-coverage effects are naturally amplified when both the training set and the model size are limited, and they provide a possible explanation for residual features that appear primarily in the distribution wings rather than as global shifts.

Finally, we test whether the generated sample reproduces the three-band color behaviour of the validation set using the same final three-band-matched catalogs. Figure~\ref{fig:color-color} compares Val and Gen in (Y--J)$-$(J--H) color space and as a function of J-band magnitude using KDE-based contours. In each panel, contours enclose fixed fractions of the probability density (10\%-90\%) and are accompanied by one-dimensional marginalized histograms. The generated sample recovers the main locus of the validation galaxies in all three projections, indicating that the model preserves the coupled photometric structure across the Y, J, and H bands rather than reproducing each band independently. In particular, the positive correlation between $(Y-J)$ and $(J-H)$ is well reproduced, and the color$-$magnitude trends with J-band brightness remain consistent between Gen and Val. The agreement is strongest in the high-density central regions of the distributions, showing that the model captures the dominant color population and its covariance structure. Residual differences are confined mainly to the outer contours, where the validation sample extends to slightly redder colors and broader wings than the generated sample. This behavior suggests that the DDPM mildly compresses the low-occupancy tails of color space, consistent with the modest tightening seen in some structural distributions, but does not introduce spurious color populations or strong shifts in the main locus. Overall, these results show that the generated galaxies reproduce the three-band photometric relations of the validation set with good fidelity.

Taken together, these stamp-level visual comparisons and distribution-level diagnostics show that the generated galaxies reproduce the main morphological and photometric properties of the \Roman-like validation sample, including their cross-band color relations.

\section{Summary and Conclusion}
\label{sec:conclusion}
This work shows that denoising diffusion probabilistic models (DDPM) provide an effective framework for generating synthetic galaxy images with morphologies representative of \Roman-like observations. The model captures a wide range of structural complexity, including isolated and blended systems, without requiring explicit supervision on these configurations. Key photometric and structural properties relevant for weak lensing analyses---AB magnitude, moment-based size, ellipticity, and peak surface brightness---are reproduced with good fidelity, indicating that the generated galaxies are statistically consistent with the validation data within the within the bootstrap-estimated sampling uncertainties.

The agreement between generated and validation samples is supported by both visual comparisons and quantitative statistical tests. In particular, distribution-level diagnostics, complemented by bootstrap resampling of the two-sample Kolmogorov–Smirnov statistic, show that show that the measured discrepancies are moderate and controlled, and are consistent with fluctuations expected from finite sample sizes. The strongest agreement is found for the peak surface-brightness distribution, while the largest deviations occur for the size proxy, where the generated sample exhibits a mild tendency toward smaller radii. These results demonstrate that the diffusion model preserves not only global morphological trends but also the dominant covariance structure among magnitude, size, ellipticity, and surface brightness, which is critical for weak lensing studies \citep{Hoekstra_2017}.

Complementary residual diagnostics show that Gen$-$Val differences are typically comparable to Train$-$Val differences, suggesting that part of the residual structure arises from finite training-set coverage and random partitioning of the parent sample rather than from systematic model failure. In addition, the generated galaxies reproduce the three-band color behavior of the validation set, indicating that the model captures the coupled photometric structure across Y, J, and H rather than matching each band independently.

Beyond realism, the proposed approach is computationally efficient, enabling the rapid generation of large numbers of high-fidelity galaxy images on standard GPU hardware. This makes diffusion-based generative models well suited for producing the extensive simulated datasets required for calibration and validation of weak lensing shape measurement pipelines. Once trained, the model can generate thousands of multi-band stamps in minutes, making it practical to scale toward survey-level simulation volumes.

Although this study is tailored to \Roman-like data, the methodology is readily extensible to other current and forthcoming cosmological surveys. Because the model learns non-parametric morphological distributions directly from imaging data, it provides a flexible alternative to purely analytic light-profile simulations and can be adapted to different instruments, pixel scales, and filter sets.

Overall, our results demonstrate that diffusion-based models can reproduce the joint structural and photometric properties of \Roman-like galaxies at a level suitable for downstream validation studies, while highlighting clear paths for further refinement---such as increasing training-set coverage or incorporating conditional generation---to reduce the remaining size-dependent discrepancies.

\section*{Acknowledgements}
The research was carried out at the Jet Propulsion Laboratory, California Institute of Technology, under a contract with the National Aeronautics and Space Administration (80NM0018D0004), © 2025. All rights reserved. In particular, this work was funded through the Jet Propulsion Laboratory’s Spontaneous Concept Research and Technology Development program, which supported this research. The High Performance Computing resources used in this work were provided by funding from the JPL Enterprise Technology, Strategy, and Cybersecurity Directorate. The authors also acknowledge the Texas Advanced Computing Center (TACC) at The University of Texas at Austin for providing computational resources that have contributed to the research results reported within this paper. Portions of this work were completed at Duke University and part of this work was also supported by the OpenUniverse effort, which is
funded by NASA under JPL Contract Task70-711320, ``Maximizing
Science Exploitation of Simulated Cosmological Survey Data Across
Surveys''. D.S. thanks Michael Troxel and Arun Kannawadi for insightful comments that improved the manuscript. This work is based on observations
made with the NASA/ESA/CSA James Webb Space Telescope. The data were
obtained from the Mikulski Archive for Space Telescopes at the Space
Telescope Science Institute, which is operated by the Association of Universities for Research in Astronomy, Inc., under NASA contract NAS 5-03127 for JWST. These observations are associated with program \#3215 and \#1963.

\vspace{0.5mm}


%

\vspace{5mm}
\facilities{All JWST JADES imaging and catalogue data used in this paper can be found in MAST:\dataset[10.17909/8tdj-8n28]{\doi{10.17909/8tdj-8n28}}} and \dataset[10.17909/fsc4-dt61]{\doi{10.17909/fsc4-dt61}}. 

\software{
Astropy \citep{2013A&A...558A..33A,2018AJ....156..123A},
Photutils \citep{photutils},
WebbPSF Toolkit \citep{Perrin2012},
roman-galaxy-ddpm \citep{roman_ddpm_zenodo}
}





\bibliography{biblio}{}
\bibliographystyle{aasjournal}


\end{document}